\def\year{2018}\relax
\documentclass[letterpaper]{article}

\usepackage{aaai18}
\usepackage{times}
\usepackage{helvet}
\usepackage{courier}
\usepackage{url}
\usepackage{graphicx}
\usepackage{color}
\usepackage{amsmath}
\usepackage{amsfonts}
\usepackage{tabularx}
\usepackage{booktabs}

\title{A Machine Learning Approach to Modeling Human Migration}

\author{Caleb Robinson \and Bistra Dilkina\\
	Georgia Institute of Technology\\
	School of Computational Science and Engineering\\
	Atlanta, Georgia. 30332\\
	\texttt{dcrobins@gatech.edu}, \texttt{bdilkina@cc.gatech.edu}
}

\nocopyright

\begin{document}
	
	\maketitle
	
	\begin{abstract}
		Human migration is a type of human mobility, where a trip involves a person moving with the intention of changing their home location. Predicting human migration as accurately as possible is important in city planning applications, international trade, spread of infectious diseases, conservation planning, and public policy development. Traditional human mobility models, such as gravity models or the more recent radiation model, predict human mobility flows based on population and distance features only. These models have been validated on commuting flows, a different type of human mobility, and are mainly used in modeling scenarios where large amounts of prior ground truth mobility data are not available. One downside of these models is that they have a fixed form and are therefore not able to capture more complicated migration dynamics. We propose machine learning models that are able to incorporate any number of exogenous features, to predict origin/destination human migration flows. Our machine learning models outperform traditional human mobility models on a variety of evaluation metrics, both in the task of predicting migrations between US counties as well as international migrations. In general, predictive machine learning models of human migration will provide a flexible base with which to model human migration under different what-if conditions, such as potential sea level rise or population growth scenarios.
	\end{abstract}
	
	\section{Introduction} \label{sec:introduction}

	Models of human mobility in their different forms are important for many reasons. Models of human commuting can help reduce traffic congestion and pollution, and can be used to drive land use policy and development choices~\cite{de2010modeling}. Models of human migration are equally important to policy makers as they can give broader estimates of how the population of an area will change in upcoming years, how labor markets might be affected~\cite{dinkelman2016long}, how infectious diseases spread~\cite{balcan2009multiscale,sorichetta2016mapping}, and how international trade will change~\cite{fagiolo2013international}. Much recent research focuses on modeling human commuting flows~\cite{masucci2013gravity,lenormand2016systematic}; however little has focused on explicitly modeling human migration.
	
	Human mobility has been traditionally modeled with the so-called gravity model, which posits that the probability of a trip between two locations decays directly as a function of the distance between them. This model was introduced in its modern form in 1946 ~\cite{zipf1946p} and has been used in many applications since~\cite{schneider1959gravity,lee1966theory,clark1980modeling,greenwood1985human,letouze2009revisiting,noulas2012tale}. More recently, the radiation model~\cite{simini2012universal} has been shown to capture long range trips better than gravity based models, and is described as `a universal model for mobility and migration patterns'. The radiation model posits that the probability of a trip will decay indirectly with distance and directly with the amount of intervening opportunities, a notion first proposed by Stouffer~\cite{stouffer1940intervening}. The radiation model has been extended several times since being proposed in 2012~\cite{masucci2013gravity,yang2014limits,ren2014predicting}. In general, gravity models have been shown to be more capable of reproducing commuting flows, i.e. human mobility at small spatial scales~\cite{lenormand2016systematic}, while radiation models have been shown to be better at reproducing migration flows, i.e. human mobility at larger spatial scales~\cite{simini2012universal}. Additionally, human migrations have been estimated by fitting generalized linear models derived from the gravity model~\cite{cohen2008international,dennett2013multilevel}. Both gravity and radiation models are analytical models  with crafted functional forms and limited input data requirements. These models are focused on explaining human migration, rely on linear relationships between independent variables, and use hand crafted features for each zone. These approaches, while useful for explaining human migration, trade predictive power for interpretability. Data sources such as the World Bank and US Census provide many zone-based features that can be algorithmically combined in a non-linear manner by tree or neural network based models to best predict human migration.
	
	\noindent Our key contributions are as follows:
	\begin{enumerate}
		\item We develop the first general machine learning formulation of the human migration prediction problem.
		\item We develop a pipeline for training machine learning models to tackle this problem that includes procedures to deal with dataset imbalance, hyperparameter tuning, and performance evaluation.
		\item We develop a custom loss function for training artificial neural networks that is more suitable for the migration prediction task.
		\item We compare the performance of machine learning models to traditional models of human migration on two datasets, and show that the machine learning models outperform the traditional models in all cases.
	\end{enumerate}
	
	\section{Traditional Migration Models} \label{sec:traditionalModels}

	In human mobility modeling, we are usually given $n$ zones with the goal of predicting the number of people, $T_{ij}$, that move from every zone $i$ to every other zone $j$.
	Traditional mobility models, such as the radiation model~\cite{simini2012universal} and the gravity model~\cite{lenormand2016systematic}, break the problem of estimating $T_{ij}$ into two pieces: estimating the total number of people $G_i$ that leave zone $i$ (also referred to as a production function), and estimating the probability $P_{ij}$ of a move occurring from $i$ to $j$. The predicted number of migrants from $i$ to $j$ would be $\hat{T}_{ij} = G_i P_{ij}$.
	As a convention, the probabilities are normalized such that, given an origin $i$, the probabilities of traveling to all other destinations sum to $1$, i.e. $\sum_{j=1}^n P_{ij} = 1$. 
	If prior information about the number of incoming and outgoing travelers per zone is known, then a constrained framework, such as the one described in Lenormand et al~\cite{lenormand2016systematic}, can be used. If this information is not known, which is the case in predicting migrations, then a \textit{production function} must be used to estimate the number of outgoing travelers per zone instead.
	A simple \textit{production function} for a dataset can be found by expressing the number of outgoing migrants of a zone as a constant fraction of the population of that zone (for counties in the US, this percentage is $\approx 0.03$~\cite{simini2012universal}). Given prior timestep's data on the population $m_i$ and the corresponding number of outgoing migrants $O_i$ in the current timestep for every zone $i$, the \textit{production function} is expressed as $G_i=M(m_i) = \alpha m_i$, where $\alpha$ is the slope of the line of best fit through the pairs ($m_i$, $O_i$).

	The traditional models of human mobility that we include in this study are: the radiation model~\cite{simini2012universal}, extended radiation model~\cite{yang2014limits}, and gravity models with both power and exponential distance decay functions~\cite{lenormand2016systematic}. The only information used by these models is: $m_i$, population of a zone $i$ for both the origin and destination zones; $d_{ij}$, the distance between two zones $i$ and $j$; and $s_{ij}$, a metric of intervening opportunities measured as the total population of all intervening zones between $i$ and $j$, defined as all zones whose centroid falls in the circle centered at $i$ with radius $d_{ij}$ (not including zones $i$ or $j$). See Table \ref{tbl:tradModels} for a description of each model.
	
%
%
%

\begin{table*}[th!]
\centering
\begin{tabular}{@{}ll@{}}
\toprule
\textbf{Model} & \textbf{Equation} \\ \midrule
Radiation & $\hat{T}_{ij} = M(m_i) \frac{m_i m_j}{(m_i + s_{ij})(m_i + m_j + s_{ij})}$ \\[0.75em]
Extended Radiation & $\hat{T}_{ij} = M(m_i) \frac{[(m_i + m_j + s_{ij})^\beta - (m_i + s_{ij})^\beta] (m_i^\beta + 1)}{[(m_i + s_{ij})^\beta + 1][(m_i + m_j + s_{ij})^\beta + 1]}$ \\[0.75em]
Gravity with Power Law & $\hat{T}_{ij} = M(m_i) \frac{m_i m_j d_{ij}^{-\beta}}{\sum_{k=1}^{n} m_k d_{ik}^{-\beta}}$ \\[0.75em]
Gravity with Exponential Law & $\hat{T}_{ij} = M(m_i) \frac{m_i m_j e^{-\beta d_{ij}}}{\sum_{k=1}^{n} m_k e^{-\beta d_{ik}}}$ \\ \bottomrule
\end{tabular}
\caption{Traditional migration models}
\label{tbl:tradModels}
\end{table*}
	
	All of these models have two parameters that must be tuned using historical data: the production function parameter, $\alpha$, that determines what fraction of the population of a zone will migrate away in a given year, and a model parameter, $\beta$. From a socio-economic point of view, traditional models have the advantage of being interpretable, however we will show that this interpretability comes at a cost of predictive accuracy, as machine learning models can use similar historic data to achieve better results.
	\section{Evaluation Methods} \label{sec:evaluationMetrics}

	To evaluate how well alternative models  perform, we use four metrics that compare how well a predicted migration matrix, $\mathbf{\hat{T}}$, recreates the ground truth values, $\mathbf{T}$.
	The first two of these ($CPC$, $CPC_d$) have been used in previous literature to evaluate human mobility models~\cite{lenormand2012universal,lenormand2016systematic}, and the other two are standard regression metrics: 
	
	\begin{description}
		\item[Common Part of Commuters ($CPC$)] This metric directly compares numbers of travelers between the predicted and ground truth matrices. It will be $0$ when the two matrices have no entries in common, and $1$ when they are identical. We note that this metric, as used in previous studies of commuting flows, is identical to the Bray-Curtis similarity score used to compare abundance data in ecological studies \cite{faith1987compositional,legendre2012numerical}.
		\begin{equation}
		CPC(\mathbf{T},\mathbf{\hat{T}}) = \frac{2 \sum_{i,j=1}^{n} min(T_{ij}, \hat{T}_{ij})}{\sum_{i,j=1}^{n} T_{ij} + \sum_{i,j=1}^{n} \hat{T}_{ij}}
		\end{equation}
		\item[Common Part of Commuters Distance Variant ($CPC_d$)] This metric measures how well a predicted migration matrix recreates trips at the same distances as the ground truth data. In this definition, $N$ is a histogram where a bin $N_k$ contains the number of migrants that travel between $2k-2$ and $2k$ kilometers. It will be $0$ when the two matrices do not have any migrations at the same distance, and $1$ when all fall within the same distances.
		\begin{equation}
		CPC_d(\mathbf{T},\mathbf{\hat{T}}) = \frac{2 \sum_{k=1}^\infty min(N_k, \hat{N}_k)}{\sum_{k=1}^\infty N_k + \sum_{k=1}^\infty \hat{N}_k}
		\end{equation}
		\item[Root mean squared error ($RMSE$)] This is a standard regression measure that will ``punish'' larger errors more than small errors. This score ranges from $0$ in a perfect match, to arbitrarily large values as the predictions become worse.
		\begin{equation}
		RMSE(\mathbf{T},\mathbf{\hat{T}}) = \sqrt{\frac{1}{n}\sum_{i,j=1}^{n} (T_{ij} - \hat{T}_{ij})^2}
		\end{equation}
		\item[Coefficient of determination ($r^2$)] This score measures the goodness of fit between a set of predictions and the ground truth values. This score ranges from $1$, in a perfect fit, to arbitrarily negative values as a fit becomes worse, and is $0$ when the predictions are equivalent to the expectation of the ground truth values.
		\begin{equation}
		r^2(\mathbf{T},\mathbf{\hat{T}}) = 1 - \frac{\sum_{i,j=1}^n (T_{ij} - \hat{T}_{ij})^2}{\sum_{i,j=1}^n (T_{ij} - \bar{T})^2}
		\end{equation}
	\end{description}
	
	In addition to the previous four metrics, we compare the ground truth number of incoming migrants and the predicted number of incoming migrants per zone using mean absolute error ($MAE$) and $r^2$. The predicted number of incoming migrants for a zone, $i$, is calculated as $\hat{v}_i = \sum_{j=1}^{n} T_{ji}$. We argue that it is important to explicitly measure how well each model performs at estimating the number of incoming migrants, because the number of incoming migrants to a location will be the most important measure for policy makers in that area. Incoming migrant predictions can inform population growth estimates and hence infrastructure planning and job analysis. 
	\section{Learning Migration Models} \label{sec:learning}

	Formally, the problem of modeling human migration is as follows: given $n$ zones, $d_1$ features describing each zone, $\mathbf{F} \in \mathbb{R}^{n \times d_1}$, and $d_2$ joint features describing features between a pair of zones, $\mathbf{J} \in \mathbb{R}^{n \times n \times d_2}$, at some timestep $t$, the objective is to predict the origin/destination \textit{migration} matrix $\mathbf{\hat{T}} \in \mathbb{N}^{n \times n}$ at the next timestep, $t+1$, where an entry $\hat{T}_{ij}$ represents the estimated number of migrants relocating from zone $i$ to zone $j$.
	Our goal is to estimate a function $f(F_{i:}, F_{j:}, J_{ij}) = \hat{T}_{ij}$, which takes the features of zone $i$ and $j$, as well as the joint features between them, as input, and directly outputs the estimated number of migrants that travel from $i$ to $j$. This approach is different from how the traditional migration models work as it does not require a \textit{production function}, but instead directly predicts $\hat{T}_{ij}$. 
	This formulation contains the simplifying assumption that migrant flows are static in time, meaning that they can be entirely determined by the features from the previous timestep. In reality migrant flows will be dependent on temporal features, such as long term developmental trends, however many places will not have enough data to take advantage of these patterns. With this formulation, our models can be applied more broadly to predict future migration patterns in locations that have only collected a single year of data. 
	
	\subsection*{Hyperparameter optimization}
	To fit $f$ for a given dataset, we will train two machine learning models, ``extreme'' gradient boosting regression (XGBoost model)~\cite{chen2016xgboost}, and a deep learning based artificial neural network model (ANN model)~\cite{lecun2015deep}. Each of these models contains several hyperparameters that must be tuned to obtain good performance on a given learning task. Our first model, the XGBoost model, is a standard machine learning model based on gradient boosting trees~\cite{mason1999boosting} that often performs very well on many regression and classification tasks. One benefit of this model is that it gives a ranking of the relative feature importances \cite{friedman2001greedy}. The parameters of the XGBoost model that we consider for hyperparameter tuning are the maximum tree depth, number of estimators, and learning rate. Our ANN model is composed of densely connected layers with ReLU activation layers\footnote{Our model is implemented in Python with the Keras library: \url{https://keras.io/}}. We tune the following ANN parameters: loss function, number of layers, layer width, number of training epochs, and training mini-batch size.
	
	\subsection*{Dealing with zero-inflated data}
	We observe that migration data is heavily zero-inflated, where in any given year, most pairs of zones do not have any migrants traveling between them, i.e. $T_{ij} = 0$ for most $(i,j)$ pairs. Considering migrations between US counties~\cite{irsMigration}, less than 1\% of the possible pairs of counties have migrations between them. This imbalance will cause problems for machine learning models. To address this problem, when creating a training dataset we undersample ``negative'' samples between pairs of zones for which there are no observed migrations. This is a na{\"i}ve technique that will necessarily throw out available information~\cite{guo2008class}. To offset this, we introduce a hyperparameter $k$ that determines the number of ``negative'' examples of migrations to train with. If there are $n_t$ pairs of zones where there are observed migrations, ``positive examples'', we include all $n_t$, and an additional $n_tk$ randomly chosen zone pairs where there are no observed migrants. This hyperparameter is included in both the XGBoost and ANN model parameter searches. We give further details on the hyperparameter tuning process in Section \ref{sec:experiments}.
	
	\subsection*{Custom ANN loss function}
	Previously we mentioned that our ANN model will consider different loss functions as a hyperparameter. Common loss functions for regression tasks include ``mean squared error'' (MSE), ``mean absolute error'' (MAE), and ``mean absolute percentage error'' (MAPE). Our preliminary experimental results show that MAE and MAPE loss functions perform poorly, partially due to their inability to punish large errors and deal with many zeros respectively. To contrast with the aforementioned zero-inflation problem, we observe that the distribution of migrant counts has a long tail, whereby few pairs of zones consistently experience large volumes of migrations. Considering these observations, and because the CPC metric is one of the key metrics of interest (described in detail in Section \ref{sec:evaluationMetrics}), we derive a new loss function based on CPC to train the ANN model with. This loss function is given as:
	\begin{equation}
	\label{eqn:cpcLoss}
	L(y,\hat{y}) = 1 - \frac{2 \sum_{i=1}^{n} min(y_{i}, \hat{y}_{i})}{\sum_{i=1}^{n} y_{i} + \sum_{i=1}^{n} \hat{y}_{i}}
	\end{equation}
Where $y_i$ is a migration flow entry from $\mathbf{T}$. The gradient update for this loss function is:
	\begin{multline}
	\label{eqn:cpcLoss2}
	\frac{\partial L(y,\hat{y})}{\partial \hat{y}_j} = \frac{2 \sum_{i=1}^n min(y_i,\hat{y}_i)}{(\sum_{i=1}^n y_i + \sum_{i=1}^n \hat{y}_i)^2} -  \frac{\begin{cases}2 & \hat{y}_j < y_j \\ 0 & \text{ otherwise}\end{cases}}{\sum_{i=1}^n y_i + \sum_{i=1}^n \hat{y}_i}
	\end{multline}
	
	Intuitively, this loss will `reward' predictions where the number of migrants matches the ground truth, while also enforcing per link absolute error to be minimized. This loss function is not exactly equivalent to the $CPC$, as during the ANN training it will only consider a single mini-batch worth of samples at a time (in our case $|y| = |\hat{y}| = |\text{mini batch}|$, where as the $CPC$ metric is a function of the entire migration matrix). Empirically this custom loss function results in better validation performance and faster training times than MSE loss.
\section{Experiments} \label{sec:experiments}

	\subsection*{Datasets}
	We perform experiments comparing the performance of traditional models to the performance of our machine learning on two datasets, the \textit{USA Migration} dataset and the \textit{Global Migration} dataset.

	The \textit{USA Migration} dataset consists of yearly intra-county migrations in the USA between 3106 counties from the IRS Tax-Stats data~\cite{irsMigration} for the 11 years in the range from 2004 to 2014. We supplement the migration data with the following 7 per-county features (taken from the US Census estimates and calculated from the Census TIGER line maps of US county boundaries): population, land area, population density, median household income, county water area, is a coastal county, and number of neighboring counties. In addition to these 7 per-county features, we add the following between-county features: distance, intervening population, intervening land area, intervening number of counties, intervening population density, and intervening median income. The intervening features are calculated based on the idea of ``intervening opportunities'' presented in the radiation model. For any given county-level variable, $x$, e.g. population, the intervening amount of that variable between counties $i$ and $j$ is defined as $s^x_{i,j}$, the sum of all $x$ in the intervening counties that fall within the circle centered at county $i$ with a radius to county $j$ (excluding $x_i$ and $x_j$).
	
	The \textit{Global Migration} dataset consists of decadal inter-country migration data between 207 countries from the World Bank Global Bilateral Migration Database ~\cite{ozden2011earth}. The \textit{Global Migration} dataset contains 5 timesteps, one every 10 years from 1960 to 2000. In the \textit{Global Migration} dataset we add the following 5 per-country features (taken from World Bank World Development Indicators data~\cite{worldbankData}): population, population density, population growth, agricultural emissions, and land area. Additionally, we include 3 between-country features: distance, intervening population, and intervening land area.
	
	For each year of data in the \textit{USA Migration} and \textit{Global Migration} datasets we create an `observation' for each pair of zones, an origin zone and destination zone (counties in the \textit{USA Migration} dataset and countries in the \textit{Global Migration} dataset). Each observation consists of the per-zone features for both the origin zone and destination zone (population of origin, population of destination, etc.) and the between-zone features of the origin and destination. This corresponds exactly to the $F_{i:}, F_{j:}, J_{ij}$ of the function $f$ (described in Section \ref{sec:learning}) that we want to learn. An observation is associated with the target number of migrants, $T_{ij}$, traveling from the origin to the destination. Formally, for a given year, $t$, number of zones, $n$, number of per-zone features, $d_1$, and number of between-zone features, $d_2$, we create a matrix of observations $\mathbf{X}_t \in \mathbb{R}^{n^2 \times (2d_1+d_2)}$ and vector of targets $Y_t \in \mathbb{R}^{n^2}$, capturing the migration flows observed in year $t+1$.
	
	\begin{table*}[ht!]
		\centering
		\resizebox{\textwidth}{!}{%
			\begin{tabular}{@{}l|rrrr|rr@{}}
				\toprule
				\textbf{USA Migrations}                         & \multicolumn{4}{c|}{\textbf{Metrics on full matrix}}                                                              & \multicolumn{2}{c}{\begin{tabular}[c]{@{}c@{}}\textbf{Metrics on incoming migrants vector}\\(Average Incoming Migrants = 3,196)\end{tabular}} \\ \midrule
				\textbf{Production Function} & \multicolumn{1}{c}{$CPC$} & \multicolumn{1}{c}{$CPC_d$} & \multicolumn{1}{c}{$RMSE$} & \multicolumn{1}{c|}{$r^2$} & \multicolumn{1}{c}{$MAE$}                                              & \multicolumn{1}{c}{$r^2$}                                             \\ \midrule
				Gravity Model Exponential Decay      & 0.53 +/- 0.01 & 0.66 +/- 0.02 & 87.4 +/- 9.0 & -1.48 +/- 0.28 & 1,216 +/- 128 & 0.67 +/- 0.03 \\
				Gravity Model Power Law Decay        & 0.56 +/- 0.01 & 0.78 +/- 0.02 & 75.7 +/- 8.0 & -0.86 +/- 0.26 & 1,129 +/- 129 & 0.72 +/- 0.04 \\
				Radiation Model                      & 0.53 +/- 0.01 & 0.76 +/- 0.02 & 47.6 +/- 5.0 & 0.26 +/- 0.09  & 1,346 +/- 148 & 0.80 +/- 0.02 \\
				Extended Radiation Model             & 0.58 +/- 0.01 & 0.83 +/- 0.01 & 35.6 +/- 3.0 & 0.59 +/- 0.03  & 1,123 +/- 117 & 0.86 +/- 0.02 \\
				XGBoost model - traditional features & 0.51 +/- 0.08 & 0.74 +/- 0.07 & 28.6 +/- 5.4 & 0.72 +/- 0.10  & 1,151 +/- 249 & 0.86 +/- 0.04 \\
				ANN model - traditional features     & 0.63 +/- 0.01 & 0.86 +/- 0.02 & 35.1 +/- 3.2 & 0.60 +/- 0.03  & 911 +/- 107   & 0.91 +/- 0.01 \\
				XGBoost model - extended features    & 0.58 +/- 0.03 & 0.78 +/- 0.02 & 24.2 +/- 1.4 & 0.81 +/- 0.02  & 968 +/- 56    & 0.89 +/- 0.02 \\
				ANN model - extended features        & 0.68 +/- 0.01 & 0.89 +/- 0.02 & 29.8 +/- 2.7 & 0.71 +/- 0.02  & 935 +/- 98    & 0.91 +/- 0.02\\ \midrule \midrule
				\textbf{No Production Function}                 & \multicolumn{1}{c}{$CPC$} & \multicolumn{1}{c}{$CPC_d$} & \multicolumn{1}{c}{$RMSE$} & \multicolumn{1}{c|}{$r^2$} & \multicolumn{1}{c}{$MAE$}                                              & \multicolumn{1}{c}{$r^2$}                                             \\ \midrule
				XGBoost model - traditional features            & 0.54 +/- 0.11             & \textbf{0.99 +/- 0.02}               & 18.5 +/- 6.1               & 0.88 +/- 0.08              & 3,091 +/- 1,740                                                        & 0.41 +/- 0.85                                                         \\
				ANN model - traditional features                & 0.63 +/- 0.02             & 0.88 +/- 0.06               & 35.3 +/- 3.5               & 0.60 +/- 0.04              & 1,188 +/- 259                                                          & 0.84 +/- 0.16                                                         \\
				XGBoost model - extended features               & 0.62 +/- 0.04             & \textbf{0.99 +/- 0.02}               & \textbf{13.0 +/- 1.5}               & \textbf{0.94 +/- 0.02}              & 2,060 +/- 622                                                          & 0.76 +/- 0.28                                                         \\
				ANN model - extended features                   & \textbf{0.69 +/- 0.01}             & 0.93 +/- 0.05               & 28.0 +/- 3.6               & 0.75 +/- 0.03              & \textbf{909 +/- 48}                                                             & \textbf{0.92 +/- 0.04}                                                         \\ \bottomrule
			\end{tabular}%
		}
		\caption{\textit{USA Migration} results. Comparison of the ANN and XGBoost models with and without a production function to traditional migration models. The values shown in the table are the average and standard deviations of the models' test performance on 2006 through 2014 data. Bold values indicate the best values per column.}
		\label{tbl:usMigrations}
	\end{table*}
	
	\begin{table*}[ht!]
		\centering
		\resizebox{\textwidth}{!}{%
			\begin{tabular}{@{}l|rrrr|rr@{}}
				\toprule
				\textbf{Global Migrations}                       & \multicolumn{4}{c|}{\textbf{Metrics on full matrix}}                                                              & \multicolumn{2}{c}{\begin{tabular}[c]{@{}c@{}}\textbf{Metrics on incoming migrants vector}\\ (Average Incoming Migrants = 674,858)\end{tabular}} \\ \midrule
				\textbf{Production Function} & \multicolumn{1}{c}{$CPC$} & \multicolumn{1}{c}{$CPC_d$} & \multicolumn{1}{c}{$RMSE$} & \multicolumn{1}{c|}{$r^2$} & \multicolumn{1}{c}{$MAE$}                                               & \multicolumn{1}{c}{$r^2$}                                              \\ \midrule
				Gravity Model Exponential Decay      & 0.16 +/- 0.00 & 0.16 +/- 0.00 & 62,218 +/- 5,341 & 0.02 +/- 0.03 & 651,194 +/- 80,220 & 0.00 +/- 0.02 \\
				Gravity Model Power Law Decay        & 0.16 +/- 0.00 & 0.15 +/- 0.00 & 61,523 +/- 5,278 & 0.05 +/- 0.00 & 628,678 +/- 79,474 & 0.03 +/- 0.03 \\
				Radiation Model                      & 0.16 +/- 0.00 & 0.14 +/- 0.00 & 62,173 +/- 5,277 & 0.02 +/- 0.00 & 614,483 +/- 79,378 & 0.04 +/- 0.02 \\
				Extended Radiation Model             & 0.16 +/- 0.00 & 0.14 +/- 0.00 & 62,108 +/- 5,299 & 0.03 +/- 0.00 & 618,576 +/- 76,150 & 0.03 +/- 0.02 \\
				XGBoost model - traditional features & 0.18 +/- 0.01 & 0.14 +/- 0.01 & 58,377 +/- 5,141 & 0.14 +/- 0.01 & 597,478 +/- 79,178 & 0.10 +/- 0.01 \\
				ANN model - traditional features     & 0.19 +/- 0.01 & 0.15 +/- 0.01 & 60,272 +/- 4,610 & 0.08 +/- 0.02 & 589,789 +/- 79,542 & 0.26 +/- 0.03 \\
				XGBoost model - extended features    & 0.21 +/- 0.01 & 0.16 +/- 0.01 & 57,909 +/- 5,409 & 0.16 +/- 0.02 & 573,090 +/- 56,987 & 0.15 +/- 0.02 \\
				ANN model - extended features        & 0.22 +/- 0.02 & 0.17 +/- 0.01 & 58,887 +/- 4,477 & 0.12 +/- 0.02 & 563,259 +/- 74,127 & 0.24 +/- 0.06 \\ \midrule \midrule
				\textbf{No Production Function}                  & \multicolumn{1}{c}{$CPC$} & \multicolumn{1}{c}{$CPC_d$} & \multicolumn{1}{c}{$RMSE$} & \multicolumn{1}{c|}{$r^2$} & \multicolumn{1}{c}{$MAE$}                                               & \multicolumn{1}{c}{$r^2$}                                              \\ \midrule
				XGBoost model – traditional features             & 0.33 +/- 0.02             & 0.59 +/- 0.03               & 52,729 +/- 5,455        & 0.26 +/- 0.26              & 938,905 +/- 172,834                                                     & 0.18 +/- 0.28                                                          \\
				ANN model – traditional features                 & 0.33 +/- 0.01             & 0.37 +/- 0.04               & 56,005 +/- 882         & 0.20 +/- 0.11              & 537,351 +/- 44,034                                                      & \textbf{0.53 +/- 0.16}                                                          \\
				XGBoost model – extended features                & \textbf{0.43 +/- 0.03}             & \textbf{0.64 +/- 0.02}               & \textbf{47,329 +/- 5,073}        & \textbf{0.42 +/- 0.13}              & 577,473 +/- 77,315                                                      & 0.48 +/- 0.34                                                          \\
				ANN model – extended features                    & 0.40 +/- 0.02             & 0.43 +/- 0.02               & 50,921 +/- 3,556        & 0.33 +/- 0.13              & \textbf{459,841 +/- 55,479}                                                      & 0.52 +/- 0.30                                                          \\ \bottomrule
			\end{tabular}%
		}
		\caption{\textit{Global Migration} results. Comparison of the ANN and XGBoost models with and without a production function to traditional migration models. The values shown in the table are the average and standard deviations of the models' test performance on 2006 through 2014 data. Bold values indicate the best values per column.}
		\label{tbl:globalMigrations}
	\end{table*}
	
	\begin{table*}[ht!]
		\centering
		\begin{tabular}{@{}lr|lr@{}}
			\toprule
			\textbf{USA Features} & \multicolumn{1}{l}{\textbf{Importance}} & \textbf{Global Features} & \multicolumn{1}{l}{\textbf{Importance}} \\ \midrule
			Intervening number of counties & 25.3\% +/- 2.4\% & Population growth of origin & 19.5\% +/- 16.0\% \\
			Population of origin (trad) & 15.7\% +/- 1.7\% & Intervening population (trad) & 12.3\% +/- 3.7\% \\
			Population of destination (trad) & 14.2\% +/- 0.9\% & Agricultural emissions of destination & 10.6\% +/- 5.8\% \\
			Intervening population (trad) & 6.1\% +/- 1.2\% & Intervening land area & 8.7\% +/- 5.6\% \\
			Is destination coastal & 4.3\% +/- 4.6\% & Population growth of destination & 7.9\% +/- 6.2\% \\
			Distance between counties (trad) & 3.7\% +/- 0.9\% & Population of destination (trad) & 6.9\% +/- 0.8\% \\
			Intervening area & 3.6\% +/- 0.9\% & Distance between counties (trad) & 6.6\% +/- 1.9\% \\
			Area of destination & 3.5\% +/- 0.5\% & Population of origin (trad) & 6.1\% +/- 1.6\% \\
			Number of neighbors destination & 3\% +/- 1.6\% & Population density of destination & 5.7\% +/- 4.5\% \\
			Water area of origin & 3\% +/- 1.3\% & Land area of origin & 5.2\% +/- 3.1\% \\ \bottomrule
		\end{tabular}%
		\caption{Top $10$ most important (extended) features in both the \textit{USA Migration} and \textit{Global Migration} datasets. The values in the table show the average and standard deviations of the information gain feature importances from an XGBoost model trained on the extended feature set for each timestep of data.}
		\label{tbl:featureImportances}
	\end{table*}
	
	\subsection*{Experimental Setup}
	
	To select the hyperparameters of the models that we described in Sections ~\ref{sec:traditionalModels} and ~\ref{sec:learning}, we consider triplets of ``years'' of data as training, validation, and testing sets. Specifically, for three years of data $\{(X_{t-2}, Y_{t-2}), (X_{t-1}, Y_{t-1}), (X_{t}, Y_{t})\}$, we call $(X_{t-2}, Y_{t-2})$ the training set, $(X_{t-1}, Y_{t-1})$ the validation set, and $(X_{t}, Y_{t})$ the test set. We tune the hyperparameters of the models using a randomized grid search with $50$ sampled hyperparameters using the training and validation sets. We select the best set of hyperparameters according to the $CPC$ score, then use those parameters to train a model on the validation set and record its performance on the test set. We repeat this process for each $(t-2, t-1, t)$ triplet of years in each dataset. Our final results are reported as averages over the test set results.
	
	A hyperparameter present in both XGBoost and ANN models is the downsampling rate, $k$. As a preprocessing step for a given year of training data $(\mathbf{X}_t, Y_t)$, we include all observations, $X_i$ where $Y_i > 0$ (let this number of samples be $m$), and choose $k*m$ random samples with replacement from the remaining observation (where $Y_i = 0$). This sampling process only takes place when training a model. When testing a model on the validation or test sets, the full datasets are always used. For experiments with the \textit{USA Migration} dataset we consider values of $k$ in a uniform distribution of integers from $5$ to $100$, while for experiments with the \textit{Global Migration} dataset we consider the uniform distribution of integers from $1$ to $5$, because the average percentage of non-zeros is $<1\%$ and $20\%$ respectively.
	
	For the XGBoost models, we sample the following parameters: maximum tree depth from $\mathcal{U}\{2,7\}$\footnote{$\mathcal{U}\{a,b\}$ is the discrete uniform distribution.}, number of estimators from $\mathcal{U}\{25, 275\}$, and learning rate uniformly in the range from $0$ to $0.5$. For the ANN models we sample the following parameters: network loss function uniformly from $\{\text{`CPC Loss'}, \text{`MSE'}\}$, number of layers uniformly from $\mathcal{U}\{1,5\}$, layer width from $\mathcal{U}\{16,128\}$, number of training epochs from $\mathcal{U}\{10,50\}$, and training mini-batch size uniformly from $\{2^9, 2^{10}, 2^{11}, 2^{12}, 2^{13}, 2^{14}\}$.

	We calibrate the parameters of the traditional models in a similar manner. Every traditional model, except for the radiation model, has two parameters, $\alpha$ and $\beta$, that must be calibrated to give useful results (the radiation model only uses $\alpha$). For each  $\{(X_{t-1}, Y_{t-1}), (X_{t}, Y_{t})\}$ pair of data (that we refer to as training and testing sets respectively), we find the value of $\alpha$ that gives the best production function on the training set. Similarly, we find the value of $\beta$ that maximizes the $CPC$ score of each traditional models on the training set. We then use these $\alpha$ and $\beta$ values to run each model on the testing set, and report the results as averages over all test set results.

    To directly compare how the ML models and traditional models perform under the same conditions, we perform experiments where the ML models are used with the same production functions as the traditional models. This imposes an artificial constraint on the ML models, as these models are able to directly estimate the number of migrations between two zones, without supplemental information on the number of outgoing migrants from each zone. To apply a production function, $M$, to the predictions made by a ML model, $\mathbf{\hat{T}}$, we create a new set of predictions, $\mathbf{\hat{T}}'$, where an entry $\hat{T}_{ij}' = M(m_i) \frac{\hat{T}_{ij}}{\sum_{k=1}^n \hat{T}_{ik}}$.

	\subsection*{Results}
	Tables \ref{tbl:usMigrations} and \ref{tbl:globalMigrations} show the average results over all years of data of the ML models and traditional models in the \textit{USA Migration} and \textit{Global Migration} datasets respectively. 
	From these tables we observe that the best traditional model for the \textit{USA Migration} dataset is the Extended Radiation model, beating the other traditional models in all metrics. None of the traditional models are able to capture the migration dynamics in the \textit{Global Migration} dataset; they all have an $r^2$ score near $0$, meaning that a model which predicts the average number of migrants for every link would perform just as well.
	The ML models perform much better.
	In the case where the ML models are constrained to the same conditions as the traditional models, using traditional features and a production function, the ANN model beats all of the traditional models in 5 out of the 6 measures in the \textit{USA Migration} datasets, and the XGBoost model beats all the traditional models in 4 out of the 6 metrics. Similarly, the ML models considerably outperform the traditional models in the \textit{Global Migration}, outperforming them in all metrics.
	Considering the extended feature set results, the ML models perform even better. The ANN and XGBoost models without a production function outperform the same models with a production function in 5 out of the 6 metrics. The XGBoost model outperforms the ANN model in 3 out of the 4 metrics that evaluate the models' per link predictions, however the ANN model performs better on the two metrics that evaluate the aggregate incoming migrant prediction performance.
	
	These results suggest that more features than those which are used by the traditional models, are needed to accurately predict human migrations. The ability of ML models to incorporate any number of additional features is one of the key motivations for using them to obtain more accurate results. Considering this, it will be insightful to understand, which of the features are most informative to the ML models. Since feature importance analysis for ANNs is quite challenging, we report in Table \ref{tbl:featureImportances} the top 10 most important features for both datasets (based on information gain) in the XGBoost model trained on the \textit{extended feature set}, averaged over all years of data. In both datasets, the intervening population feature is in the top 4 important features which validates the intuition that intervening opportunities are important in predicting migrations. The most important feature in the \textit{USA Migration} dataset is the number of intervening counties between two locations, a simpler form of the intervening population idea. In the \textit{Global Migration} dataset, the population growth of the origin is the most important feature on average, with a large standard deviation. In some years this feature is very important, however in other years it is less so. Intuitively, population growth will be correlated with the amount of incoming migration. During relatively stable years, with small population growth, other features will be more predictive of migration.
	
\begin{figure*}[ht!]
		\centering
		\includegraphics[width=0.49\linewidth]{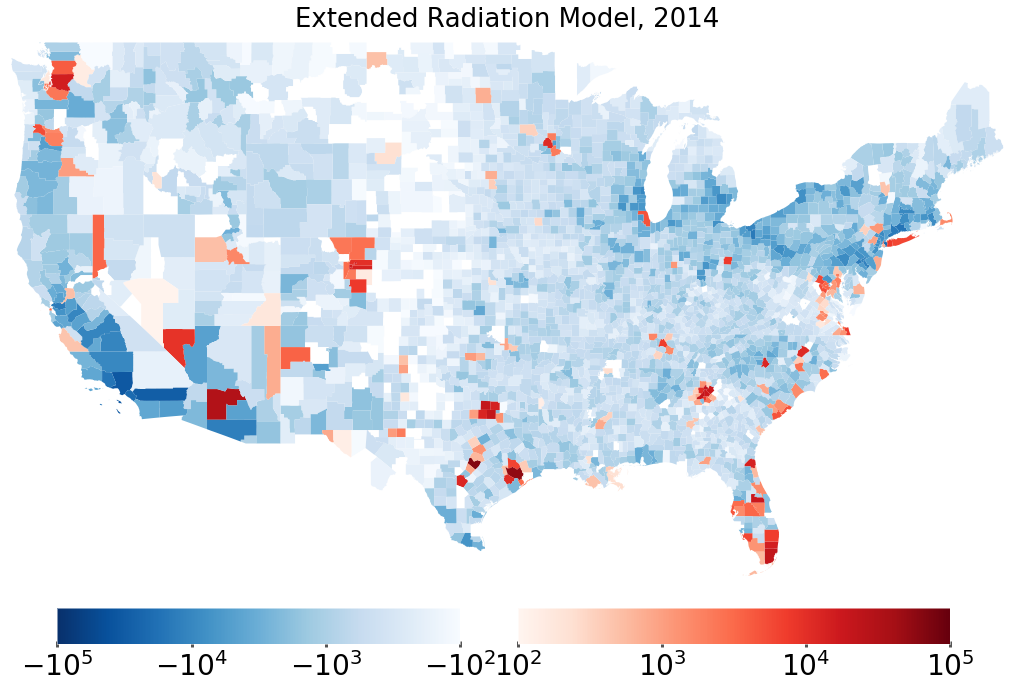}
        \includegraphics[width=0.49\linewidth]{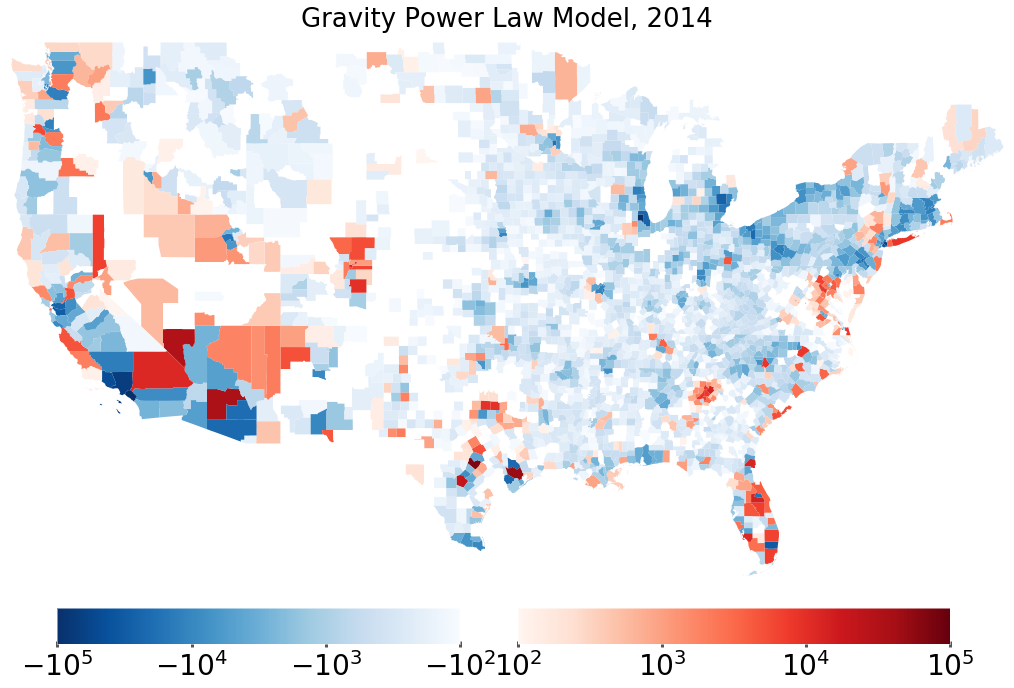} \vspace{-1cm} \\
		\includegraphics[width=0.49\linewidth]{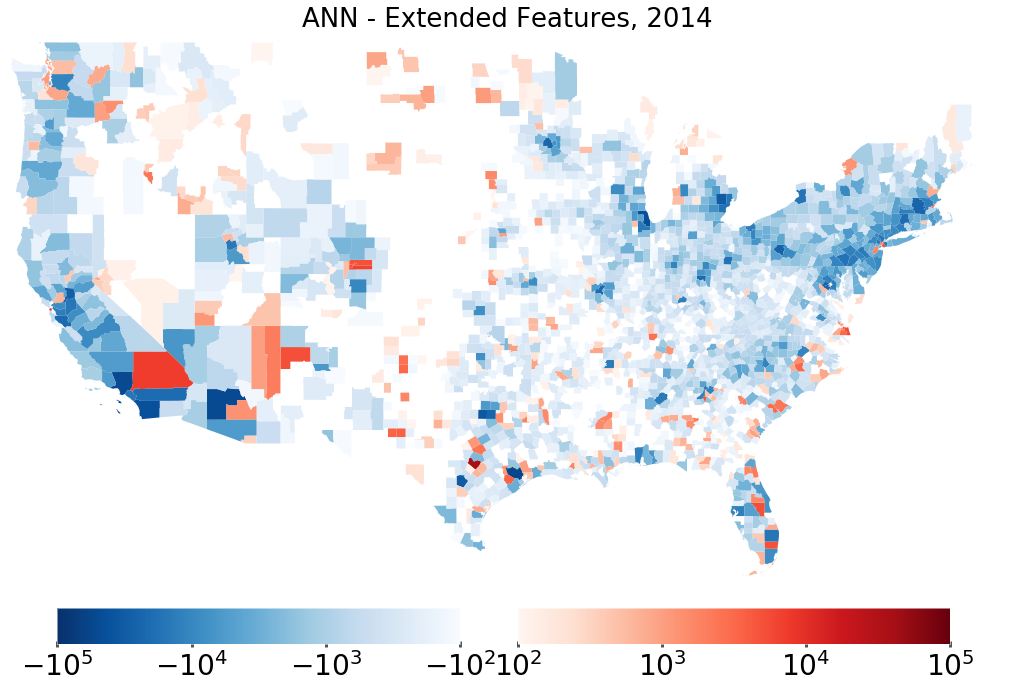}
		\includegraphics[width=0.49\linewidth]{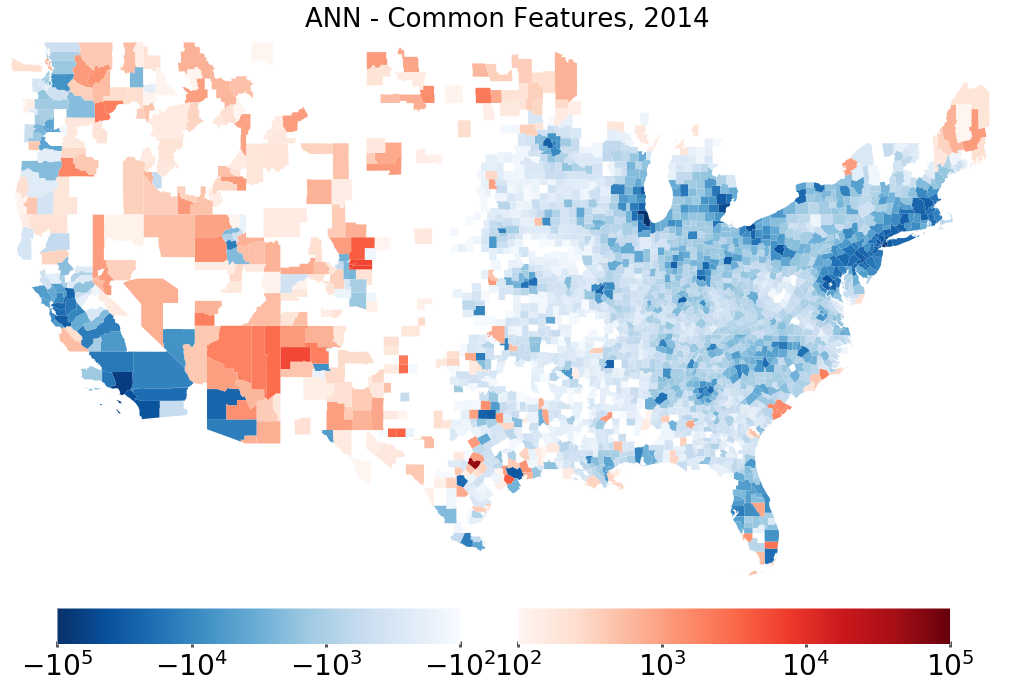} \vspace{-1cm} \\
        \includegraphics[width=0.49\linewidth]{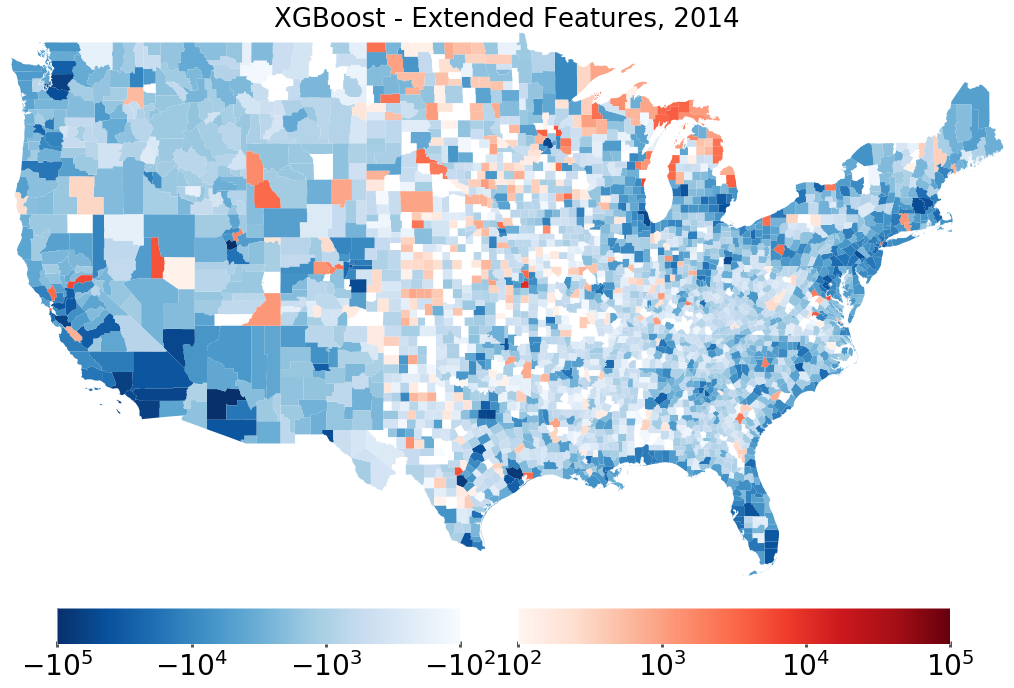}
        \includegraphics[width=0.49\linewidth]{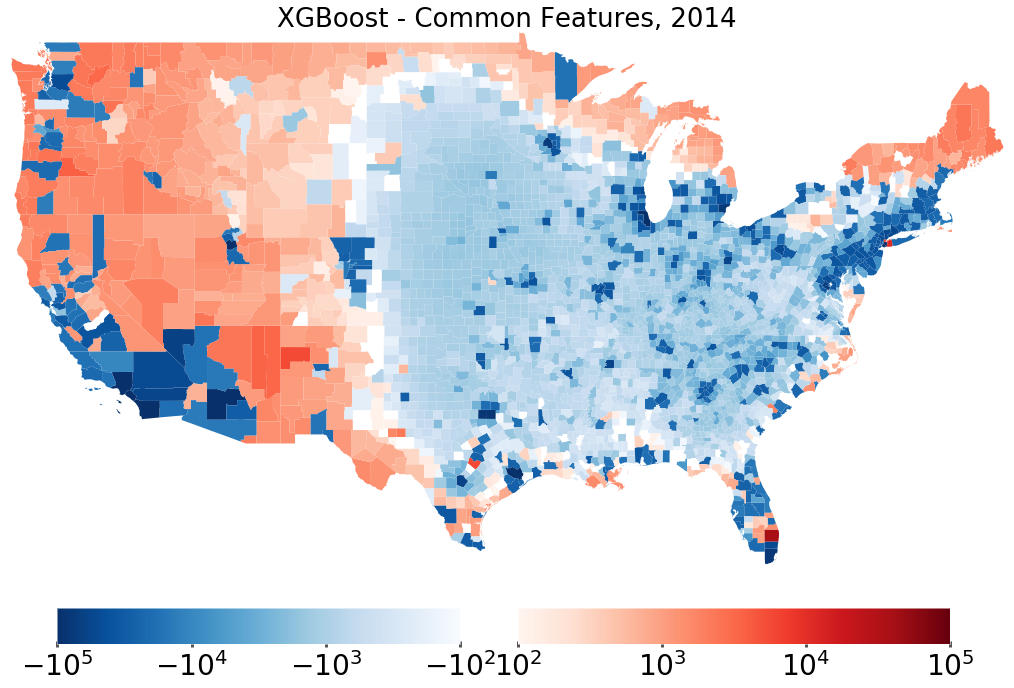}
		\caption{\textbf{\textit{USA Migrations} modeling error.} These maps show the difference between the ground truth number of incoming migrants and predicted number of incoming migrants per county for 6 models in 2014. Blue corresponds to overestimation by the model, red to underestimation by the model, and white if the model accurately predicts the correct number of incoming migrants. \textbf{Top row} shows the results for the Extended Radiation model and Gravity model with power law distance decay. \textbf{Middle row} shows the results for ANN models trained with the extended and common feature sets. \textbf{Bottom row} shows the results for XGBoost models trained with the extended and common feature sets.}
		\label{fig:errorMaps}
	\end{figure*}

	
	In Figure \ref{fig:errorMaps} we show the difference between the actual and predicted numbers of incoming migrants per county for the two best traditional models, and all of the ML models without production functions. From these maps we can see that between ML models, those trained with the extended feature set perform better than those trained with only the traditional features. Specifically, without the extended features, the ML models underpredict the number of migrations to the western portion of the United States. When the extended features are taken into account, the models are able to correct for this spatial bias. Holding with the experimental results, we can see that the ANN model with extended features best captures the incoming migrant distributions per county. The ANN model is able to more accurately match the number of migrants that travel to rural areas (e.g. to the midwestern US), compared to the traditional models that consistently over estimate the numbers of migrants to rural areas. In general, these maps agree with our empirical results, that the ANN model (with the lowest average incoming migrants $MAE$) is able to best predict migrations.

	\section{Conclusion}
	
	With the increasing availability of high resolution socio-economic data in countries that also record human migrant flows, it is possible to use machine learning models of human migration rather than traditional gravity or radiation models. Machine learning models offer greater levels of modeling flexibility, as they can combine many input features in non-linear ways that can not be captured by static equations. Furthermore, machine learning models can be easily customized to the problem or country at hand.
	
	We develop two machine learning based models for the task of predicting human migration flows, for both between counties in the US and between countries across the world. We compare these models to traditional human migration models using two sets of features and show that our models outperform the traditional models in most of the evaluation metrics.
	
	We would like to extend this work to better explain human migration through a more complete analysis of features included in the model, and through incorporating different models. While the XGBoost model can provide a ranking of feature importances, this does not fully explain the dynamics that drive human migration. Additionally we would like to study how these migration models could be specialized to predict migrations under extreme weather events. Hurricanes and other natural disasters can displace large populations, and determining where these populations will resettle would provide an unique planning tool for policy makers. To achieve these goals, higher resolution migration data, on both spatial and temporal scales, will need to be obtained. Extreme weather events, by definition, are short lived and their effects will be better estimated and predicted at local scales.
	
	\bibliographystyle{aaai}
	\bibliography{citations} 
	
\end{document}